\documentclass[conference]{IEEEtran}

\usepackage{ifpdf}

\usepackage{cite}

\usepackage[cmex10]{amsmath}

\usepackage{array}

\usepackage[tight,footnotesize]{subfigure}

\usepackage{graphicx}
\usepackage{here}
\usepackage{siunitx}
\usepackage[utf8]{inputenc}
\usepackage{booktabs}
\usepackage{url}
\usepackage{algorithm}
\usepackage{algpseudocode}
\usepackage{wasysym}
\usepackage{nicefrac}

\begin{document}
	%
	\title{Feasibility Study of Financial P2P Energy Trading in a Grid-tied Power Network}

	\author{\IEEEauthorblockN{M Imran Azim, S. A. Pourmousavi, Wayes Tushar, and Tapan K. Saha}
		\IEEEauthorblockA{School of Information Technology and Electrical Engineering, \\
			The University of Queensland,
			Brisbane, QLD 4072, Australia \\
			E-mail: m.azim@uq.net.au, a.pour@uq.edu.au, w.tushar@uq.edu.au, saha@itee.uq.edu.au \\
		}
		
	}
	\maketitle

	\begin{abstract}
		\boldmath
		This paper studies the applicability of peer-to-peer (P2P) energy trading in a grid-tied network. The main objectives are to understand the impact of the financial P2P energy trading on the network operation, and thus demonstrate the importance of taking various issues related to power network into account while designing a practical P2P trading scheme. To do so, a simple mechanism is developed for energy trading among prosumers without considering any network constraints, as done by many existing studies. Once the trading parameters, such as the energy traded by each prosumer in the P2P market and the price per unit of energy are determined, the developed scheme is tested on a low-voltage (LV) network model to check its feasibility of deployment in a real P2P network. It is shown that although the considered trading scheme is economically beneficial to the participating prosumers compared to the existing incentive mechanisms (such as feed-in-tariff), it could be unfit for real deployment due to violating bus voltage limits during multiple P2P trading executed simultaneously. Further, the grid operator may experience financial losses for compensating the losses during P2P transactions. 
		
	\end{abstract}
	

	\section{Introduction}
	\label{sec:introduction}
	
	Rooftop solar installation has been beneficial for the society due to the numerous economic and environmental advantages that it demonstrates \cite{rooftopsolar}. It has also allowed traditional passive consumers to become active prosumers. These prosumers can locally generate clean energy to meet their own demand and sell the energy surplus for monetary gains \cite{miller2017social}. Feed-in-tariff (FiT) incentive scheme permits prosumers to sell their excess energy to the grid (or equivalently to the retailers) at a fixed price per unit of energy \cite{ye2017analysis}. In order to promote the rapid uptake of rooftop solar generation and to benefit the local community, FiT tariff was initially set high, for instance, around 44 \cent/kWh in Queensland, Australia in 2008 \cite{QLDsolar}. It continued for few years and then dropped considerably to 8 c/kWh in 2014 due to raising electricity retail prices for non-solar households. In fact, some applicants started installing rooftop solar beyond their household requirements with the intention of earning money via FiT, which was against the purpose of the FiT scheme \cite{QLDsolar}. Currently, Queensland has fixed FiT rate nearly to 11 \cent/kWh \cite{QLDfit}, while the average electricity retail price is 20.3 \cent/kWh (during off-peak hours) and 25.6 \cent/kWh (during peak hours) \cite{QLDeprices}. Consequently, the benefit to a prosumer for participating in FiT scheme has become very marginal. Therefore, alternative consumer-centric market approaches are required to consider in which significant number of prosumers can receive financial benefits through local energy trading.
	
	Peer-to-peer (P2P) energy trading is such type of distributed market structure that enables prosumers to trade energy among themselves at negotiated prices. As a result, they are likely to receive more attractive economic benefits than the present FiT scheme \cite{tushar2018transforming}. 
	The operation of the P2P trading can be divided into two parts, namely financial trading and physical trading \cite{tushar2018transforming}. A number of proposals has been presented by the researchers on advanced financial trading framework. For instance, prospective P2P market price schemes, policies, rules and regulations have been studied in \cite{long2017peer} and \cite{mengelkamp2018designing}. Cooperative trading approach has been proposed in \cite{tushar2018peer} to incentivise prosumers. In this framework, prosumers form groups depending on their cooperative behaviour and the profit earned via P2P trading is equally distributed among the prosumers of the group. An alternative approach is used for P2P trading in \cite{zhang2018peer}, in which an individual prosumer can make financial trading decisions independently.
	
	
	In the existing literature, however, a little attention has been paid towards the physical P2P energy transfer. It is obvious that a valid P2P transaction should not violate the bus voltage limits set by the upper grid. In addition, the impact of P2P trading on the network losses in a grid-connected system should be identified clearly. For example, losses allocation during local energy trading has been provided in \cite{zizzo2018technical} and \cite{nikolaidis2018graph}. However, they did not show the financial impact of the transaction losses on the grid. For the voltage, authors in \cite{guerrero2018decentralized} have shown that their proposed financial model does not violate the network voltage limits during small amount of energy transfer. In reality, it is most likely that small amount of energy exchange at a single feeder would not create notable voltage problems. Nevertheless, significant presence of rooftop solar generators, stationary and mobile storage devices, and modern energy management systems will increase the amount of locally-traded energy in the future, which may result in violating the bus voltage limit.
	
	This paper focuses on investigating the network problems that a financially attractive P2P trading can introduce in a power network. To achieve this goal, a simple P2P trading mechanism is presented which can find out the P2P market solutions quickly without requiring rigorous computational efforts. Then, the monetary gains that P2P trading can offer to the prosumers compared to the existing mechanisms (i.e., retail prices and FiT rate) are assessed. Finally, the trading quantity is applied on a LV network to identify the possible network issues associated with P2P trading at the physical layer. 
	
	Please note that the financial layer in this paper (i.e., the trading algorithm) is intentionally designed to be simple and intuitive because its simplicity allowed us to explain the physical layer's problems more effectively, without distracting the readers by the complexity of the financial trading algorithm.
	
	
	The rest of the paper is organised as follows: Section~\ref{sec:financialP2Ptrading} describes a simple rule-based P2P trading mechanism. The issues that a P2P trading is likely to create on the physical network along with intuitive methods to compute them are discussed in Section~\ref{sec:physicalP2Ptrading}. Section~\ref{sec:simulationstudies} contains the simulation studies and results. Finally, the concluding remarks are given in Section~\ref{sec:conclusion}.

	\section{Financial P2P Trading Algorithm}
	\label{sec:financialP2Ptrading}
	
	In this section, prosumers' selection process, P2P pricing strategy, and the trading algorithm are explained.
	
	\subsection{Market Setup}
	\label{subsec:marketsetup}
	
\begin{figure} [h]
		\centering
		\includegraphics [width=7.4cm]{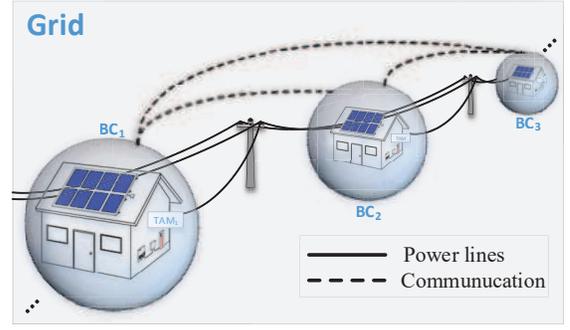}
		\caption{A local market structure based on P2P communication.}
		\label{fig:Fig1}
	\end{figure}

	It is assumed that all prosumers are equipped with a blockchain (BC) account, by which they can communicate with other participants directly. In short, BC is a decentralised communication technology that offers fast, efficient, secure, reliable, and transparent P2P transactions \cite{tushar2018transforming}. They are also assumed to have a smart meter known as trans-active meter (TAM). TAM simultaneously monitors and records the local generation at the prosumers' side. The TAM will send a signal to the corresponding BC account when there is an energy surplus or deficiency. Once signal is received by the BC platform, prosumers will start communicating with each other \cite{tushar2018transforming}. Fig.~\ref{fig:Fig1} illustrates a hypothetical P2P market and the direct communication links among prosumers using BC technology. It is assumed that the prosumers are not allowed to sell and buy energy simultaneously. Therefore, prosumers can be identified either as a seller or a buyer for a given time interval. It is also assumed that the sellers use energy internally as much as needed and only offer excess energy to the P2P market. In addition, the sellers and buyers will declare their prices in every P2P intervals. Once agreed on a BC ledger for a trading interval, the P2P market rules, presented in \cite{mengelkamp2018designing}, will not permit the prosumers to change their prices for that interval. 
	
	In this paper, a prospective approach is adopted to arrange the priority order of sellers and buyers when multiple prosumers wish to participate in the P2P trading. The sellers' order are settled based on the least declared price. For the buyers, priority is given in the order in which they registered using their BC account, i.e., BC registration order (BCRO). 
	
	\subsection{P2P Pricing Mechanism}
	\label{subsec:P2Ppricingmechanism}
	
	Let $\mathcal{S}_t=\{\mathbf{c}_s,\mathbf{p}_s\}$ and $\mathcal{B}_t=\{\mathbf{c}_b,\mathbf{p}_b\}$ be the set of price and quantity pairs of all sellers and buyers, respectively, registered in the market for trading period $t$. The sellers price and quantity are $\mathbf{c}_s$ in $\{c_{s,1},c_{s,2},\dots,c_{s,S}\}$ and $\mathbf{p}_s$ in $\{p_{s,1},p_{s,2},\dots,p_{s,S}\}$, respectively, for $S$ sellers. The buyers price and quantity are $\mathbf{c}_b$ in $\{c_{b,1},c_{b,2},\dots,c_{b,B}\}$ and $\mathbf{p}_b$ in $\{p_{b,1},p_{b,2},\dots,p_{b,B}\}$, respectively, for $B$ buyers. Also assume that the sets are sorted based on the approach proposed in Subsection~\ref{subsec:marketsetup}. Based on the mid-market rate (MMR) method \cite{long2017peer}, the P2P trading price $\lambda\left(k\vert i,j\right)$ for $\rho\left(k\vert i,j\right)$ kW power exchange between seller $i$ and buyer $j$ can be defined as the middle value of $c_{s,i}$ and $c_{b,j}$ for a specific transaction:
	\begin{equation}
	\label{eq:tradingprice}
	\lambda\left(k\vert i,j\right)=\frac {c_{s,i}+c_{b,j} }{2}
	\end{equation}
	where $\lambda\left(k\vert i,j\right)$ is the price of the $k$\textsuperscript{th} transaction in $\Lambda_t=\{\lambda\left(1\vert \cdot,\cdot\right),\lambda\left(2\vert \cdot,\cdot\right),\cdots,\lambda\left(K\vert \cdot,\cdot\right)\}$ for $\rho\left(k\vert i,j\right)$ power exchange in $\mathcal{P}_t=\{\rho\left(1\vert\cdot,\cdot\right),\rho\left(2\vert\cdot,\cdot\right),\cdots,\rho\left(K\vert\cdot,\cdot\right)\}$ at time $t$. Please note that the participants can trade with multiple prosumers in the market at time $t$. Therefore, the number of transactions, $K$, is not necessarily the same as the number of buyers and/or sellers. In that case, average P2P trading prices for each seller and buyer, $\overline{\lambda}_i$ and $\overline{\lambda}_j$, are calculated as follows:
	\begin{equation}
	\begin{cases}
	\overline{\lambda}_i=\nicefrac{\sum\limits_{j}\lambda\left(\cdot\vert i,j\right)}{J_i}\qquad \forall \,i\in\{1,\cdots,S\} \\
	\overline{\lambda}_j=\nicefrac{\sum\limits_{i}\lambda\left(\cdot\vert i,j\right)}{I_j}\qquad \forall \,j\in\{1,\cdots,B\}
	\end{cases}
	\label{eq:avgprice}
	\end{equation}
	
	\noindent where $J_i$ and $I_j$ are the number of transactions for seller $i$ and buyer $j$, respectively. In reality, both local generation and demand vary in time. Accordingly, the prices offered by the buyers and sellers vary over time. When local generation is not available or not sufficient to cover the load demand, the buyers should pay at the grid price, which is higher than the local generation price. In the existing arrangement at the distribution level in many countries (such as Australia), sellers inject excess energy to the network in exchange for the FiT price,  $\lambda_{\textrm{FiT}}$. Any deficiency in the load demand will also be compensated by the grid (or retailers) at the retail price, $\lambda_{\textrm{On--Peak}}$ and $\lambda_{\textrm{Off--Peak}}$ for time-of-use (ToU) tariffs. Hence, sellers' profit ($\mathcal{M}_i$) and buyers' savings ($\mathcal{M}_j$) from participating in the P2P market for the entire day can be calculated by,
	\begin{equation}
	\begin{cases}
	\mathcal{M}_i=\sum\limits_t\sum\limits_j\left(\lambda\right(\cdot\vert i,j\left)-\lambda_{\textrm{FiT}}\right)\times\rho\left(\cdot\vert i,j\right)\times \Delta_t \\
	\mathcal{M}_j=\sum\limits_t\sum\limits_i\left(\lambda_{(\cdot)}-\lambda\right(\cdot\vert i,j\left)\right)\times\rho\left(\cdot\vert i,j\right)\times \Delta_t
	\end{cases}
	\label{eq:benefits}
	\end{equation}
	
	\noindent where $\Delta_t$ is the trading time interval in hour, and $\lambda_{(\cdot)}$ represents ToU prices of the grid, i.e., $\lambda_{\textrm{On--Peak}}$ and $\lambda_{\textrm{Off--Peak}}$, in \cent/kWh. In Subsection~\ref{subsec:economicbenefits}, daily profits and savings are calculated for a case study with P2P trading framework. 
	
	When a deal is finalised between seller $i$ and buyer $j$, the agreed quantity, $\rho\left(k\vert i,j\right)$, will be reported to the grid operator. Also, the actual amount of power traded during that time interval will be recorded by the TAMs, and the measurements will be available to the grid for billing. At the end of a billing cycle, P2P transactions will be deducted by the electricity retailers from the local generation and demand to calculate sellers' payment and buyers' bill. The TAM readings can also be used to settle the payments among P2P participants. 

	\subsection{Trading Algorithm}
	\label{subsec:tradingalgorithm}
	
	The trading mechanism is given in Algorithm~\ref{alg:P2Palgorithm}. Unlike other approaches, such as game theoretic \cite{tushar2018peer, zhang2018peer}, it finds the solution quickly while requiring less computational efforts. As explained earlier, sellers and buyers are sorted based on the minimum seller price and the buyer registration order in $\mathcal{S}_t$ and $\mathcal{B}_t$ sets, respectively, in the first step. In the next step, sellers' power, $\mathbf{p}_s$, and buyers' demand, $\mathbf{p}_b$, at time $t$ are adjusted so that the total supply and load demand are equal, i.e., $\sum\mathbf{p}_s=\sum\mathbf{p}_b$. The algorithm informs affected prosumers about the generation curtailment or energy shortage within the market. Nevertheless, any mismatch will be compensated by the grid, and it will be settled outside of the P2P market at the retailer FiT or ToU prices, as explained in subsection~\ref{subsec:P2Ppricingmechanism}. In the final step, trading among sellers and buyers is carried out at P2P prices. The procedure starts with the first seller, $\left(c_{s,1},p_{s,1}\right)$, and buyer, $\left(c_{b,1},p_{b,1}\right)$, from the list. The trading price, $\lambda\left(1\vert 1,1\right)$, will be calculated by Eq.~\eqref{eq:tradingprice} and the traded power will be $\rho\left(1\vert 1,1\right)$. If the buyer's load, $p_{b,1}$, is less than the seller's supply, $p_{s,1}$, the remainder energy will be traded with the next buyer in the list, which forms transaction $\rho\left(2\vert 1,2\right)$ at $\lambda\left(2\vert 1,2\right)$ price. Similarly, if the seller is not able to satisfy the load, $p_{b,1}$, next seller will be considered, i.e., transaction $\rho\left(2\vert 2,1\right)$ at $\lambda\left(2\vert 2,1\right)$ price. This process will continue until all buyers and sellers are satisfied. 
	
	\begin{algorithm}[h]
		\caption{A Rule-based P2P Trading Mechanism}
		\label{alg:P2Palgorithm}
		\begin{algorithmic}[1]
			
			\State \textbf{Input Parameters:} $ \mathcal{S},  \mathcal{B} $ 
			
			\For {$S$ sellers and $B$ buyers}
			\State \textsc{\textbf{Step 1}: Sorting Prosumers}
			\State \hspace{\algorithmicindent}$ \mathcal{S} \gets \textrm{min}$\ $\left\lbrace\mathbf{c}_{s}\right\rbrace $
			
			\State \hspace{\algorithmicindent}$ \mathcal{B} \gets \textrm{BCRO}$ 
			
			\State \textsc{\textbf{Step 2}: Power Balance}
			\While {$(\sum\mathbf{p}_s\neq\sum\mathbf{p}_b)$}
			\If {$(\sum\mathbf{p}_s>\sum\mathbf{p}_b)$}
			\State Curtail most expensive seller
			\Else
			\State Curtail last registered buyer
			\EndIf
			\EndWhile

			\State \textsc{\textbf{Step 3}: P2P Trading}
			
			\For{$j$ in $B$ buyers}
			\While{$\sum_i\rho\left(\cdot\vert i,j\right)\neq p_{b,j}$}
			\State purchase from cheapest seller
			\State calculate price in Eq. \eqref{eq:tradingprice}
			\State calculate average price in Eq. \eqref{eq:avgprice}
			\EndWhile
			\EndFor
			\EndFor
			
		\end{algorithmic}
	\end{algorithm}
	
	\section{Physical P2P Trading}
	\label{sec:physicalP2Ptrading}
	
	Most of the P2P trading markets are expected to be developed in grid-tied systems. 
	To examine the impact of P2P trading on the grid-connected networks, two case studies are presented, which are explained in details below:

	\subsection{Bus Voltage Assessment}
	\label{subsec:busvoltageassessment}
	
	Whether participating in the P2P market or selling at the FiT price to the grid, the excess local generation injected to the network can cause over-voltage issues in the low-voltage (LV) networks. In an attempt to reduce the voltage issues, modern rooftop solar inverters are manufactured to sink/source reactive power to regulate voltage at the point of common coupling (PCC) to some extent. According to some grid codes, advanced inverters are recommended for the new rooftop solar installations. In this case, it is important to consider the impact of active power trading on the network voltage profiles so that the amount of traded active power is decided based on the network conditions. Otherwise, the inverters automatically curtail active power injection to regulate voltage at the PCC. Let's consider a scenario, in which seller $i$ trades $\sum_j\rho(k\vert i,j)$ kW power at time $t$ with buyers. If injecting that amount of power triggers upper voltage violation, the inverter will curtail injected power (by absorbing reactive power) for voltage regulation. As a result, seller $i$ will be penalised in the final billing settlement for not delivering $\sum_j\rho(k\vert i,j)$ kW power. This problem has been shown in the simulation results for a specific case in Subsection~\ref{subsec:highvoltageissue}.  
	
	\subsection{Hypothetical Transaction Losses}
	\label{subsec:networklossassessment}
	
	In many energy trading algorithms, the losses caused by the P2P transactions are avoided without adequate technical justifications. As a result, buyers and sellers agree upon a quantity and price that does not account for the losses in the system, which means that the buyers never receive the exact amount of power promised by the sellers due to the losses (if any). In fact, the further the buyers are from the sellers, the more could be the transaction losses. In the case of connection to an upper grid, the transaction losses (if any) should be compensated by the grid without being reflected in the electricity bills, i.e., free of charge. The reason is that the TAM readings only verifies the injected power at the seller's PCC, it does not give a hint about the incurred transaction losses (if any). To calculate the hypothetical losses for one or more P2P transactions, the below steps are followed:
	
	\textbf{Step 1}: Run a power flow study, where all injected power are assumed to be zero, i.e., $\mathbf{p}_s=0$, which is called Case I. Then, measure the total power imported from the grid, $\mathcal{P}_g^{\textrm{I}}$;
	
	\textbf{Step 2}: Perform a power flow study with the hypothetical P2P market solutions, which is termed as Case II. Then, reads the power flow through the distribution grid and call it $\mathcal{P}_g^{\textrm{II}}$.
	
	The difference between $\mathcal{P}_g^{\textrm{II}}$ and $\mathcal{P}_g^{\textrm{I}}$ is the total losses associated with all P2P transactions within a given time slot.

	\section{Simulation Studies}
	\label{sec:simulationstudies}
	
\begin{figure} [h]
		\centering
		\includegraphics[width=8.5cm]{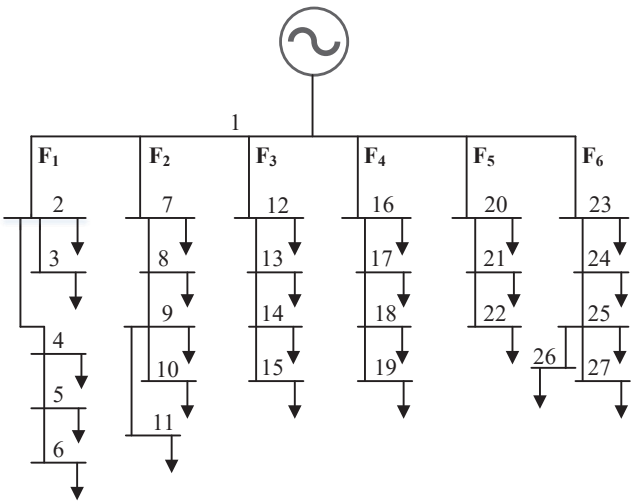}
		\vspace{-0.1 in}
		\caption{0.4 kV Danish LV distribution grid model \cite{ahmadi2016assessing}.}
		\label{fig:Fig2}
	\end{figure}

	This section considers a 27 bus distribution test system, shown in Fig.~\ref{fig:Fig2}, for a comprehensive simulation study. Bus, branch, and load data of the system are taken from \cite{ahmadi2016assessing}. All the sellers and buyers are connected to different buses randomly. Three simulation studies are performed pertaining to P2P trading. First study discusses the economic benefits that both sellers and buyers can receive by participating in the P2P market in a typical day, whereas the other two studies highlight the network issues that are likely to happen during physical P2P trading.
	
	\subsection{Economic Benefits}
	\label{subsec:economicbenefits}
	
	In this simulation study, five prosumers have been considered, the details of which are given in Table \ref{tab:TableI}. At the beginning of each trading interval, prosumers have been allowed to submit their preferred prices arbitrarily within a specified price range. The price range has been set below the retailer offers and above the FiT rate so that all participants make profits. Without loss of generality, the prosumers' prices are assumed to be the same in all trading intervals. 
	Importantly, it is not necessary that sellers' price should be lower than buyers' price or vice-versa. Rather, MMR method has been used to calculate the trading price from sellers' and buyers' declared prices.


	\vspace{-0.18 in}
	\begin{table} [h]
		\centering
		\caption{Details of the Prosumers}
		\vspace{-0.1 in}
		\label{tab:TableI}
		\begin{tabular}{ |c|c|c|c| }
			\hline
			Prosumers & Trading Identity & Declared  Price (\cent/kWh) & BCRO \\
			\hline
			1 & Buyer 1 & 15 & 2\\
			\hline
			2 & Buyer 2 & 14 & 1\\
			\hline
			3 & Seller 1 & 17 & 4\\
			\hline
			4 & Seller 2 & 15 & 5\\
			\hline
			5 & Seller 3 & 16 & 3 \\
			\hline
		\end{tabular}
	\end{table}
	\vspace{-0.1 in}
	
\begin{figure} [h]
		\centering
		\includegraphics [width=8cm]{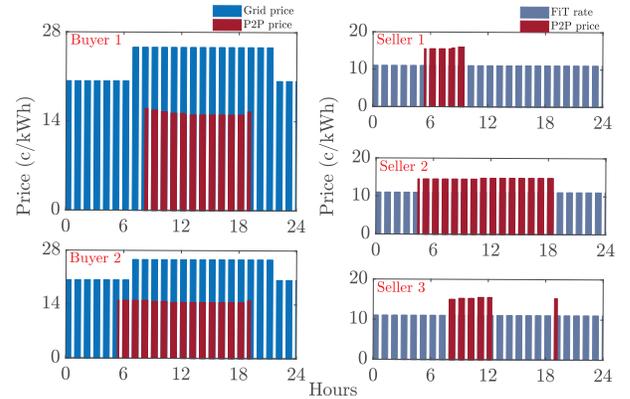}
		\vspace{-0.1 in}
		\caption{Financial benefits of P2P trading for all participants.}
		\label{fig:Fig3}
	\end{figure}

	The daily energy profile of the prosumers with 15 minutes interval as well as the local retail prices are taken from a Brisbane-based research centre. It is assumed that the net load (internal demand minus local generation) represents sellers' excess energy. Therefore, the negative value of the net load enables sellers to exchange energy with the buyers. The rule-based P2P trading algorithm is executed for the entire day for the prosumers at hand. Fig.~\ref{fig:Fig3} shows the P2P trading prices for the participants compared to the retail prices. Since the grid prices, i.e., $\lambda_\textrm{On-Peak}$ and $\lambda_\textrm{Off-Peak}$, are quite higher than the P2P trading prices, i.e., $\lambda\left(\cdot\vert i,j\right)$, buyers are expected to save some money. In this study, Buyer 2 and Buyer 1 can save \$3.13 and \$1.5, respectively, in the course of a day. Please note that Buyer 2 registered first and got the chance to trade first in the P2P market at lower prices. On the other hand, the FiT rate, $\lambda_\textrm{FiT}$, is by far lower than the P2P trading prices. Hence, sellers can earn more profits by trading in the P2P market compared to selling energy to the grid at FiT rate. In the simulation study, Seller 2 earns \$1.24 more at the end of the day, followed by Seller 3 (\$0.3) and Seller 1 (\cent8.8). Please note that Seller 1 asked for higher price and thus, traded less in the P2P market. In summary, it can be concluded that the more prosumers participate in the P2P trading, the more they can gain financial benefits.
	\vspace{-0.1 in}
	\vspace{-0.18 in}
	
	\subsection{High Voltage Issue}
	\label{subsec:highvoltageissue}
	
	Let's assume a scenario, where different houses at buses 2-6 on feeder $F_{1}$, in Fig.~\ref{fig:Fig2}, have installed rooftop PV panels. Seller 2 (at bus 2) sells 2.8 kW to Buyer 2 (at bus 3) and 1.4 kW to Buyer 1 (at bus 6) at time $t$. Seller 1 (at bus 5) and Seller 3 (at bus 4) also start selling their excess energy at the same time. They are injecting 3.7 kW and 3.5 kW power, respectively, to trade with buyers at bus 14 and bus 19. Therefore, $\mathbf{p}_s=\{3.7, 4.2, 3.5\}$ kW and $\mathbf{p}_b=\{2.8, 1.4, 3.7, 3.5\}$ kW.
	
\begin{figure} [h]
		\centering
		\includegraphics [width=8.8cm]{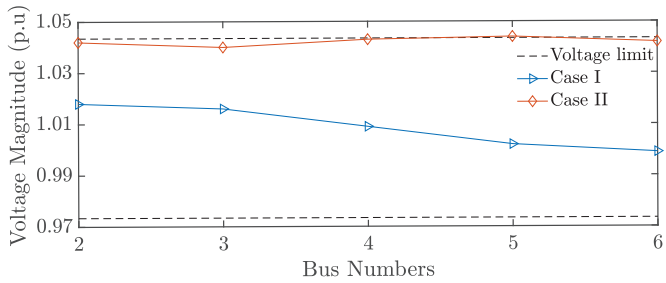}
		\vspace{-0.23 in}
		\caption{Bus voltages at $F_{1}$ feeder during multiple P2P trading.}
		\label{fig:Fig4}
	\end{figure}

	Fig.~\ref{fig:Fig4} reveals that simultaneous P2P transactions can rise the bus voltages of $F_{1}$ beyond the limit prescribed in Queensland, Australia \cite{QLDvoltage}. For instance, voltage at bus 5 crosses the upper voltage limit while buses 2, 4 and 6 are very close to exceed it as well. Therefore, multiple P2P trading inside a single feeder can cause over-voltage in the network. Consequently, as described in \ref{subsec:busvoltageassessment}, injected power will be curtailed for voltage regulation and the financial analysis, done by the Algorithm~\ref{alg:P2Palgorithm}, is going to become obsolete for practical implementation.
	
	\vspace{-0.14 in}
	
	\subsection{Losses Compensation Issue}
	\label{subsec:networklosscompensationissue}
	
	Let's consider another scenario in which a buyer at bus 27 purchases 5 kW from a seller at bus 26, that results in 0.02 kW hypothetical losses. If the same buyer purchases the same amount of power from a seller at far-off bus 6, the transaction losses increase to 0.48 kW.
	Thus, if the cost of losses is paid by the buyer, the sellers at near distance with higher prices might become more appealing than the sellers at a long distance offering a lower price. 
	Therefore, it is important to consider the impact of losses within the financial trading layer. 
	
	In a grid-tied network, if the transaction losses are not paid by the P2P participants, grid has to compensate the losses for free. 
	Let's assume the trading scenario for a particular time interval, given in Table \ref{tab:TableII}, where eight simultaneous transactions are finalised. As is demonstrated in Table \ref{tab:TableII}, the total P2P losses is 1.21 kW (4.1\% of the total exchanged power) for all transactions, which is high for such a small LV network. In fact, if similar P2P transactions are carried out throughout the year, the grid has to supply 42.4 MW excess power in that year to compensate P2P losses which can cause approximately \$2200-\$2700 financial losses.
	
	\vspace{-0.18 in}
	\begin{table} [h]
		\centering
		\caption{Determination of P2P Transaction Losses}
		\vspace{-0.1 in}
		\label{tab:TableII}
		\begin{tabular}{ |c|c|c| }
			\hline
			P2P Trading  & Exchanged Power  & Grid's Losses Compensation  \\
			Buses, $\left(i,j\right)$   &  (kW), $\rho\left(\cdot\vert i,j\right)$ & (kW) \\
			\hline
			(2,3) & 2.8 & \\
			\cline{1-2}
			(2,6) & 1.4 & \\
			\cline{1-2}
			(4,19) & 3.5 & \\
			\cline{1-2}
			(5,14) & 3.7 & \\
			\cline{1-2}
			(6,27) & 5.0 & 1.21\\
			\cline{1-2}
			(8,21) & 4.0 & \\
			\cline{1-2}
			(10,25) & 4.7 & \\
			\cline{1-2}
			(11,26) & 4.9 & \\
			\hline
		\end{tabular}
	\end{table}
	\vspace{-0.1 in}

	For a real LV P2P network with thousands of buses sprawled over a large territory, the transaction losses could be more pronounced. It can be intensified by the fact that anyone from anywhere in the network can join a P2P network to trade energy. Therefore, the amount of losses that has to be compensated by the grid in a real system with significant P2P transactions cannot be ignored. Otherwise, the grid operator may experience substantial financial losses annually.

	\section{Conclusion}
	\label{sec:conclusion}
	This paper highlights the possible network issues associated with the P2P energy trading market. To do so, a simple P2P trading framework is considered and implemented in a real test system. Then, several simulation studies are carried out to reveal the benefits of trading in such market, and concerns regarding bus voltages and hypothetical P2P transaction losses. 
	It is shown that both buyers and sellers can benefit from a local P2P market. However, the simulation results figure out that over-voltage can happen when the rooftop solar penetration is relatively high within an LV feeder during multiple P2P transactions. Moreover, transaction losses might be a real problem in such trading mechanism, where losses could be compensated by the grid without financial return. 
	
	Please note that the model presented in this paper to evaluate the feasibility of P2P trading mechanism can be applied to any other financial algorithms without any modifications.   
	
	In future research, new strategies will be developed to bring power network issues into the P2P energy trading framework.

	\bibliographystyle{ieeetr}

\begin{thebibliography}{10}

\bibitem{rooftopsolar}
``Solar panels for home.''
  \url{https://www.greenlivingtips.com/articles/advantages-of-rooftop-solar-power.html}.

\bibitem{miller2017social}
W.~Miller and M.~Senadeera, ``Social transition from energy consumers to
  prosumers: Rethinking the purpose and functionality of eco-feedback
  technologies,'' {\em Sustainable Cities and Society}, vol.~35, pp.~615--625,
  Nov. 2017.

\bibitem{ye2017analysis}
L.-C. Ye, J.~F. Rodrigues, and H.~X. Lin, ``Analysis of feed-in tariff policies
  for solar photovoltaic in {China} 2011--2016,'' {\em Applied Energy},
  vol.~203, pp.~496--505, Oct. 2017.

\bibitem{QLDsolar}
``Queensland solar bonus scheme policy guide,'' {\em Department of Natural
  Resources, Mines and Energy, Queensland Government, Australia, Report}, 2018.

\bibitem{QLDfit}
``Feed-in-tariff information for {Queensland}.''
  \url{https://www.solarquotes.com.au/systems/feed-in-tariffs/qld/}.

\bibitem{QLDeprices}
``Electricity prices in {South East Queensland}.''
  \url{https://www.business.qld.gov.au/running-business/energy-business/energy-pricing/electricity-prices}.

\bibitem{tushar2018transforming}
W.~Tushar, C.~Yuen, H.~Mohsenian-Rad, T.~Saha, H.~V. Poor, and K.~L. Wood,
  ``Transforming energy networks via peer to peer energy trading: Potential of
  game theoretic approaches,'' {\em IEEE Signal Processing Magazine}, vol.~35,
  no.~4, pp.~90--111, Jul. 2018.

\bibitem{long2017peer}
C.~Long, J.~Wu, C.~Zhang, L.~Thomas, M.~Cheng, and N.~Jenkins, ``Peer-to-peer
  energy trading in a community microgrid,'' in {\em Proc. of the IEEE Power \&
  Energy Society General Meeting}, pp.~1--5, IEEE, Jul. 2017.

\bibitem{mengelkamp2018designing}
E.~Mengelkamp, J.~G{\"a}rttner, K.~Rock, S.~Kessler, L.~Orsini, and
  C.~Weinhardt, ``Designing microgrid energy markets: A case study: The
  {Brooklyn} microgrid,'' {\em Applied Energy}, vol.~210, pp.~870--880, Feb.
  2018.

\bibitem{tushar2018peer}
W.~Tushar, T.~K. Saha, C.~Yuen, P.~Liddell, R.~Bean, and H.~V. Poor,
  ``Peer-to-peer energy trading with sustainable user participation: A game
  theoretic approach,'' {\em IEEE Access (Early Access), doi:
  10.1109/ACCESS.2018.2875405.}, 2018.

\bibitem{zhang2018peer}
C.~Zhang, J.~Wu, Y.~Zhou, M.~Cheng, and C.~Long, ``Peer-to-peer energy trading
  in a microgrid,'' {\em Applied Energy}, vol.~220, pp.~1--12, Jun. 2018.

\bibitem{zizzo2018technical}
G.~Zizzo, E.~R. Sanseverino, M.~G. Ippolito, M.~L. Di~Silvestre, and P.~Gallo,
  ``A technical approach to {P2P} energy transactions in microgrids,'' {\em
  IEEE Transactions on Industrial Informatics (Early Access), doi:
  10.1109/TII.2018.2806357}, 2018.

\bibitem{nikolaidis2018graph}
A.~Nikolaidis, C.~A. Charalambous, and P.~Mancarella, ``A graph-based loss
  allocation framework for transactive energy markets in unbalanced radial
  distribution networks,'' {\em IEEE Transactions on Power Systems (Early
  Access), doi: 10.1109/TPWRS.2018.2832164}, 2018.

\bibitem{guerrero2018decentralized}
J.~Guerrero, A.~C. Chapman, and G.~Verbi{\v{c}}, ``Decentralized {P2P} energy
  trading under network constraints in a low-voltage network,'' {\em IEEE
  Transactions on Smart Grid (Early Access), doi: 10.1109/TSG.2018.2878445},
  2018.

\bibitem{ahmadi2016assessing}
R.~Ahmadi~Kordkheili, S.~A. Pourmousavi, M.~Savaghebi, J.~M. Guerrero, and
  M.~H. Nehrir, ``Assessing the potential of plug-in electric vehicles in
  active distribution networks,'' {\em Energies}, vol.~9, no.~1, pp.~34--50,
  Jan. 2016.

\bibitem{QLDvoltage}
``New statutory voltage limits for {Queensland}.''
  \url{https://www.dnrme.qld.gov.au/energy/initiatives/statutory-voltage-limits},
  2018.

\end{thebibliography}

\end{document}